\documentclass[aps,prl,reprint]{revtex4-2}

\usepackage{amsmath,amssymb,amsfonts,color,graphicx,tabularx}

\usepackage[unicode=true,colorlinks=true]{hyperref}

\usepackage{bm}

\usepackage{subfigure}

\usepackage{tikz}

\hypersetup{linkcolor=blue,citecolor=blue,urlcolor=black}

\usepackage{etoolbox}
\patchcmd{\thebibliography}{\section*{\refname}}{}{}{}

\begin{document}

\title{Inter-band optical transitions of helical Majorana edge modes in topological superconductors}

\author{Han Bi$ ^1 $}
\author{James Jun He$ ^{2 } $}
\email{jun\_he@ustc.edu.cn}
\affiliation{
	$ ^1 $International Center for Quantum Design of Functional Materials (ICQD), 
	Hefei National Research Center for Interdisciplinary Sciences at the Microscale, 
	University of Science and Technology of China, Hefei, Anhui 230026, China \\
	$ ^2 $Hefei National Laboratory, Hefei, Anhui 230088, China
}
\date{\today}

\begin{abstract}
  The search for evidence of Majorana states on the edges of topological superconductors (TSCs) is challenging due to the difficulty of detecting such charge-neutral electronic quasiparticles. Local microwave spectroscopy has been shown to be a possible method to detect propagating Majorana modes,  where a spatially focused light beam must be used. Here, we show that helical Majorana modes in TSCs allow inter-band transitions and thus contribute to optical conductivity under a spatially uniform light. The existence of such a signal requires the system to break certain symmetries so that the projection of the charge current operator onto helical Majorana edge states leads to inter-band hybridization terms. The general form of this contribution under a tunable time-reversal breaking field is derived, which is valid in the sub-gap low-frequency regime where the edge energy spectrum is linear, and numerical results are obtained in three TSC models, showing remarkable consistency with the analytical prediction. In comparison, the current operator for normal helical edge states, such as in quantum spin Hall insulators, does not cause inter-band transitions and the related optical conductivity vanishes unless the time-reversal symmetry is broken. Our results may help guide feasible experiments to provide evidence of Majorana edge modes in TSCs. 
\end{abstract}

\maketitle

\section{Introduction}

The search for Majorana modes in condensed matter physics \cite{kitaev2001unpaired,wilczek2009majorana, AliceaRPP2012, Leijnse2012, BeenakkerARCMP2013, ElliottRMP2015,PradaNRP2020,FlensbergNRM2021,JackNRP2021} has been a critical and challenging problem.  Such quasiparticles are believed to exist in topological superconductors (TSCs) where they may show up as one-dimensional (1D) propagating modes as well as zero-dimensional bound sates. 

Various systems have been predicted to host propagating helical \cite{FuPRL2008,Qi2009PRL,liu2011helical,deng2012majorana,deng2013multiband,WangPRB2014,YangPRB2015,sun2016helical,parhizgar2017highly,haim2019time,ZhangRXPRL2019,ZhangRXPRL2019b,ZhangRXPRL2021} or chiral \cite{read2000paired,fu2009probing,akhmerov2009electrically,qi2010chiral,daido2017majorana,wang2018multiple,lian2018topological,he2019platform,fu2021chiral} Majorana modes with or without time-reversal symmetry, respectively. There has been experimental results \cite{kasahara2018majorana,banerjee2018observation,WangZY2020} consistent with propagating-Majorana scenarios, but conclusive evidence of such exotic states in TSCs is yet to be achieved. The difficulties not only exist in experimental techniques, but also in theoretical principles to interpret or predict the Majorana signals. Despite possessing nonzero velocity, propagating Majorana modes do not carry any charge because their particles and anti-particles are identical \cite{kitaev2001unpaired}. This neutrality makes the detection of Majorana edge modes in TSCs much harder than normal edge states. 

Strictly speaking, however, the neutrality is true only on average since Majorana modes alone do not preserve the U(1) gauge symmetry. This makes it possible for them to couple with external electromagnetic field, leading to particular optical responses that may serve as evidence of chiral Majorana modes \cite{he2021prl,he2021local}. In this case, translation symmetry needs to be strongly broken to see the predicted optical signal due to the absence of vertical optical transition. As a result, a highly focused light beam is required.

Here, we show that, without breaking the transnational symmetry along the edge, the microwave absorption of helical Majorana modes (HMMs) in time-reversal invariant TSCs is nonzero  and may be used as an effective detection method. We begin with a generic discussion with an effective theory containing only the edge states. The current operator, assumed to be determined by the current operator of the bulk TSC system projected onto the edge, is directly written down by physical arguments at this stage. A general form of the optical absorption is obtained. Then we investigate this phenomenon in several  models of time-reversal invariant TSCs including $p \pm ip$-wave superconductors, topological insulator (TI) thin films in proximity to superconductivity, and doped quantum spin Hall (QSH) insulators. We show that they all share the same features in the sub-gap low-energy region where the optical transitions happen among the Majorana edge states.

\section{Effective edge theory}

\begin{figure}
	\includegraphics[width=0.5\textwidth]{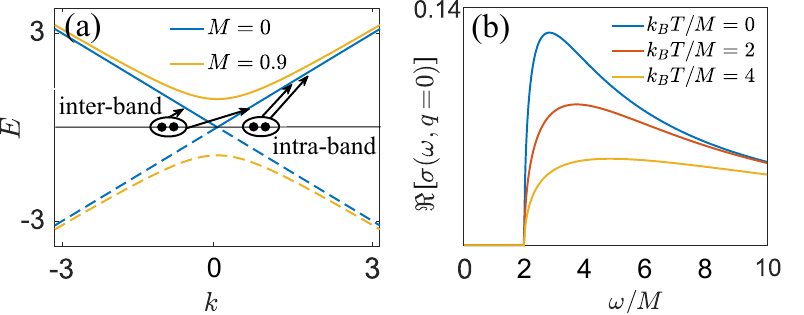}
	\caption{(a) The energy spectra of the helical Majorana modes without ($ M=0 $) and with ($ M\neq 0 $) time-reversal symmetry. Intra-band and inter-band  optical transitions are schematically shown during which a Cooper pair is broken into two Majorana modes. (b) The real-part homogeneous optical conductivity, $\Re[\sigma(\omega,0)]$, as a function of the frequency $\omega $ given by Eq. (\ref{eq:4}). }
	\label{fig:1}
\end{figure}

A minimum theory of the helical Majorana edge states of a time-reversal invariant TSC may be described by the following Hamiltonian,
\begin{equation}
    \mathcal{H}=\sum_{-k_0<k<k_0}\Gamma_k^\dagger (vk\sigma_3+M\sigma_2)\Gamma_k,
    \label{eq:1}
\end{equation}
where $\Gamma_k=[\gamma_{1k},\gamma_{2k}]$ denotes the edge HMMs, $\sigma_{1,2,3}$ are the Pauli matrices, $k_0$ is the momentum cutoff,
and $M$ is a time-reversal breaking term that opens a gap in the 1D spectrum. When $M=0$, one readily find that the time-reversal symmetry $\mathcal{T}=i\sigma_2 \mathcal{K}$ and the particle-hole symmetry $\mathcal{P}=\mathcal{K}$ are both preserved. When $M\neq 0$, the energy eigenvalues are $\pm \xi_k$ with $\xi_k=\sqrt{(vk)^2+M^2}$, which has a gap of $M$, as shown in Fig. \ref{fig:1}(a). 

The corresponding current density operator $j(x)$ is more conveniently written down in the real space. Considering its sign flip under both $\mathcal{T}$ and $\mathcal{P}$, applying the Majorana algebra $\{\gamma_{i}(x),\gamma_{j}(x^{'})\}=\delta_{ij}\delta(x-x^{'})$, and keeping up to the first order spatial derivative, one obtains the only possible form 
\begin{equation}
	j(x)= -ia(\gamma_1\partial_x \gamma_1+\gamma_2 \partial_x\gamma_2)-ib\gamma_1\gamma_2,
	\label{eq:2}
\end{equation}
where $a$ and $b$ are real.

The current operator in the momentum space is given by Fourier transformation $j_q=\int  j(x) e^{-iqx} dx$ and the optical conductivity is given by the Kubo formula,
\begin{equation}
   \Re[\sigma(\omega,q)]=\frac{1}{\omega L}\Im\int_0^{k_BT} d\tau e^{i\omega_n\tau}\langle T_\tau j_{-q} (\tau)j_q(0)\rangle.
   \label{eq:3}
\end{equation}
where $\Re $ $(\Im)$ denote the real (imaginary) part.
For a uniform detecting light we only need to consider the $q=0$ component. Eq. (\ref{eq:3}) is calculated in the eigenstate basis, yielding 
\begin{equation}
    \Re[\sigma(\omega,0)]=\frac{b^2}{4v\omega^2}{\sqrt{\omega^2-(2M)^2}}\tanh{\frac{\omega}{2k_BT}}.
    \label{eq:4}
\end{equation}
When the temperature $T\rightarrow 0$ and $M=0$, the $\Re[\sigma(\omega,0)]$ curve decreases as $1/\omega$. For the time-reversal breaking case, the optical response is zero inside the edge gap, i.e. when $\omega<2M$ which is the lowest energy to break a Cooper pair into two Majorana states with the same energy and opposite momenta. $\Re[\sigma(\omega,0)]$ reaches the maximum value at $\omega=2\sqrt{2}M$ and decreases with further increasing $\omega$, as shown in Fig. \ref{fig:1}(b).
Generally, the momentum cutoff $k_0$ in Eq. (\ref{eq:1}) originates from a energy cutoff $\Delta=vk_0$ which corresponds to the topological gap of a TSC. This gap is usually smaller than the SC gap and is often of the order of $0.1$ meV, corresponding to microwave. 

Note that the right-hand side of Eq. (\ref{eq:4}) vanishes if $b=0$ and thus the last term in Eq. (\ref{eq:2}) is crucial and responsible for the inter-band transitions. This term is taken for granted up to now. In the following, we study concrete TSC models where the inter-band term of $j(x)$ appears by  breaking the inversion symmetry. 

\section{$p\pm ip$ superconductors}
As the simplest case, let us first consider a $p\pm ip$-wave TSC described by the following Hamiltonian
\begin{align}
    H_{p\pm ip}&=(\frac{\hbar^2 k^2}{2m}-\mu)\tau_3\sigma_0+A_p(k_x\tau_0\sigma_1+k_y\tau_3\sigma_2)\\
    &+\Delta_p (k_x\tau_1\sigma_0- k_y\tau_2\sigma_3)
    \label{eq:9}
\end{align}
with the basis $\Psi^\dagger(\bm{k})=[\psi_{\bm{k}\uparrow}^\dagger,\psi_{\bm{k}\downarrow}^\dagger,\psi_{-\bm{k}\uparrow},\psi_{-\bm{k}\downarrow}]$. Here $\bm{\tau}$ and $\sigma_i$ are Pauli matrices ($\sigma_0$ being the identity matrix) acting on the particle-hole and spin degrees of freedom respectively. A pair of gapless helical Majorana states appear at the boundary of the 2D system and are protected by the time-reversal symmetry $\mathcal{T}$, which can be broken by a small Zeeman term $H_z=B_p \tau_3\sigma_1$. The spin orbit coupling (SOC) term proportional to $A_p$ is necessary in order to induce optical response to a uniform light, because the Hamiltonian (\ref{eq:9}) with $A_p=0$ commutes with the spin operator $\sigma_3$ and there is no mixing between $\uparrow$ and $\downarrow$, indicating $b=0$ in Eq. (\ref{eq:2}) and thus vanishing vertical transition. 


The current operator along the $\hat{\bm{x}}$ direction is given by 
\begin{equation}
    j_x=\begin{pmatrix}
         j_n(\bm{k}) & \\ & -j_n^T(-\bm{k})
    \end{pmatrix}
\end{equation}
where $j_n(\bm{k})=e \hbar k_x/m \sigma_0+e A_p\sigma_1$ is the $k$-derivative of the normal state Hamiltonian $H_n=(\hbar^2k^2/2m-\mu)\sigma_0+B_p \sigma_1+A_p\bm{k}\cdot\bm{\sigma}$ multiplied by the electron charge $e$. The Kubo formula calculated in the energy eigenstate basis leads to
\begin{align}
    \sigma(\omega)=\frac{i\hbar}{\Omega}\sum_{k_x,m,n}&\frac{\vert\langle nk_x\vert j_x\vert mk_x\rangle\vert^2}{\xi_{mk_x}-\xi_{nk_x}}\notag \\
    &\times\frac{f(\xi_{nk_x})-f(\xi_{mk_x})}{\hbar\omega+\xi_{nk_x}-\xi_{mk_x}+i\eta},
\end{align}
where $\Omega$ stands for the area of the shining light and $f(\epsilon)$ represents the Fermi distribution function. 
Note that, in the absence of SOC effect, $j_x$ is proportional to the identical matrix and no inter-band transition occurs due to the orthogonality of the eigenstates, $\vert nk\rangle$ and $\vert mk\rangle$. Adding SOC terms breaks the inversion symmetry and produces spin-mixing terms in the current operator. Then, we expect nontrivial inter-band transition described by the effective 1D theory in the previous section.  

\begin{figure}
    \centering
    \subfigure{\includegraphics[scale=0.90]{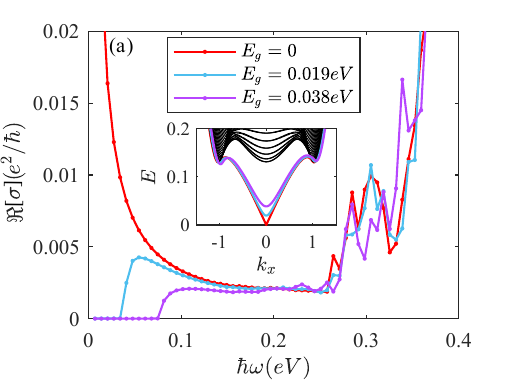}}\vspace{-5mm}
    \subfigure{\includegraphics[scale=0.95]{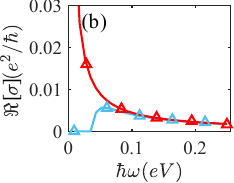}}\hspace{-1mm}
    \subfigure{\includegraphics[scale=0.95]{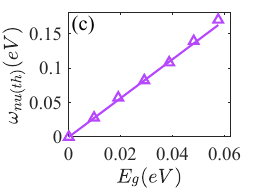}}
    \caption{(a) The frequency dependence of optical conductivity for various values of the edge gap $E_g$. The common parameters: pairing amplitude $\Delta=0.2$, SOC strength $A_p=0.05$, mass term $m=1$, chemical potential $\mu=0.5$ and $k_BT=10^{-4}$. $k_B$ is the Boltzmann constant. The inset shows the corresponding energy spectra. (b) The numerical (dots) and the analytical (lines, obtained with Eq. (\ref{eq:4})) of the real-part conductivity for both the time-reversal invariant (red) and the time-reversal broken case (blue). (c) The corresponding peak position $\omega_{nu(th)}$ where the real-part conductivity reaches its maximum under various values of $E_g$. }
    \label{fig:2}
\end{figure}

Figure \ref{fig:2}(a) shows the real-part conductivity $\Re[\sigma(\omega)]$ obtained numerically. When the Zeeman field $B_p=0$,  the optical absorption induced by the in-gap HMMs diverges at as $ \omega\rightarrow 0 $. This is in agreement with the prediction of the effective theory in which $\Re[\sigma(\omega)]\sim 1/\omega$. 
For  nonzero but small $B_p$, the edge states acquire a gap, denoted by $ E_g $. Thus, there is no absorption for $ \omega<2E_g $ unless thermal excitation due to finite temperature is considered. 
$\Re[\sigma]$ rapidly increases near $\omega>2E_g$ and reaches a maximum at $\omega_{nu(th)}$. It decreases as $ \omega $ further goes up until it reaches the bulk gap, where the contribution of the bulk states dominates. 
Figure \ref{fig:2}(b) shows the real-part optical conductivity contributed by the edge states, with or without time-reversal symmetry, together with the corresponding analytical results of the effective 1D theory.  
Fig.  \ref{fig:2}(c) shows the positions  of the maximum for various values of the energy gap. 
They both show great agreement between the numerical and the analytical results.


\begin{figure}[h]
    \centering
    \includegraphics[scale=0.88]{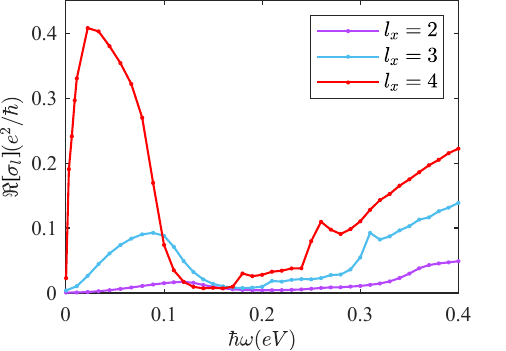}\vspace{-1mm}
    \caption{The frequency dependence of the real-part local conductivity $\Re[\sigma_l(\omega)]$ of the $p\pm ip$-wave superconductors system. The results for different values of the shining length $l_x$ is plotted in different colors. Other related parameters are set to be the same to those in Fig. \ref{fig:2} with $E_g=0$.}
    \label{fig:3l}
\end{figure}
The optical conductivity of the $p\pm ip$-wave superconductor in response to a locally distributed detecting light is also studied by transforming Eq.(\ref{eq:9}) into a tight-binding model. The current density along $\hat{\bm{x}}$ direction is
$
    j_x(\bm{r})=e[\frac{i \hbar }{2m}\psi_{\bm{r}}^\dagger\psi_{\bm{r}+\bm{\hat{{x}}}}+A_p\psi_{\bm{r}}^\dagger\psi_{\bm{r}}+h.c.]
$
The size of detecting area is determined by the limits $1\leq x\leq l_x$ and $1\leq y\leq l_y$, where $l_y$ is chosen to cover the spread of the edge state wave functions. The   current operator to calculate the optical response  in the detecting area is given by
\begin{equation}
    J_x=\frac{1}{l_x}\sum_{m=1}^{l_x}\sum_{n=1}^{l_y}j_x(\bm{r}+m\bm{\hat{x}}+n\bm{\hat{{y}}}).
\end{equation}
Figure \ref{fig:3l} shows the results for different shining lengths $l_x$ obtained using the recursive Green's function method. For small $l_x$ we get similar shape of the $\Re[\sigma_l(\omega)]$ curve compared to the chiral case \cite{he2021prl}. A major difference is the non-zero value of $\Re[\sigma_l(\omega=0)$, which is  a consequence of the inter-band term ($ \sim b $) in the current operator of Eq. (\ref{eq:2}). 
As $l_x$ increases, the peak shift towards lower frequency and $\Re[\sigma_l(\omega)]$ and the peak height increases. For very large $l_x$, the result becomes similar to the uniform case, as expected since the limit $ l_x \rightarrow \infty  $ recovers uniformity. 
 
\section{TI thin film}
The surface states of TIs can be used to design a time-reversal invariant TSC by proximity to conventional superconductors. 
If the SC order parameters induced on the two surfaces of a TI thin film are different by  a  phase $ \pi $, a pair of HMMs appear on the edges \cite{liu2011helical,parhizgar2017highly}. 
With only the surface states  considered, the normal state Hamiltonian can be written as
\begin{equation}
    H_0^{TF}(\bm{k})=2A_t\tau_z \bm{d}\cdot\bm{\sigma}+m_{\bm{k}}\tau_x\sigma_0
    \label{eq:TI}
\end{equation}
under the basis  $\Psi_{\bm{k}}^\dagger=[c_{\bm{k},+,\uparrow}^\dagger,c_{\bm{k},+,\downarrow}^\dagger,c_{\bm{k},-,\uparrow}^\dagger,c_{\bm{k},-,\downarrow}^\dagger]$ where $ \pm $ denotes the two surfaces. The vector $\bm{d}=[k_x, k_y,0]$ and the function $m_{\bm{k}}=m_0-t_f(k_x^2+k_y^2)$. The first term of Eq. (\ref{eq:TI}) describes the two Dirac cones located at the two surfaces and the second term represents the inter-surface coupling. 
An  inversion-symmetry-breaking term (originating from the substrate, for example) is needed to induce vertical optical transition \cite{mattis1958theory,ahn2021theory}, which may be, for example, a Rashba SOC on one surface,
\begin{equation}
    H_R^{TF}(\bm{k})=\alpha (\tau_0+\tau_z)\bm{d}\times\bm{\sigma}.
    \label{eq:TIb}
\end{equation}
Equations (\ref{eq:TI})-(\ref{eq:TIb}) added by an s-wave pairing, $\Delta(\bm{k})=i\Delta\tau_z\sigma_y$, form the total Hamiltonian. The sign difference of the order parameter between the upper and lower surfaces guarantees the time-reversal symmetry, which can be slightly broken by an external Zeeman field $H_z=B\tau_0\sigma_z$ or by a deviation from the exact $\pi$-phase difference.

\begin{figure}
    \centering
    \includegraphics[scale=0.85]{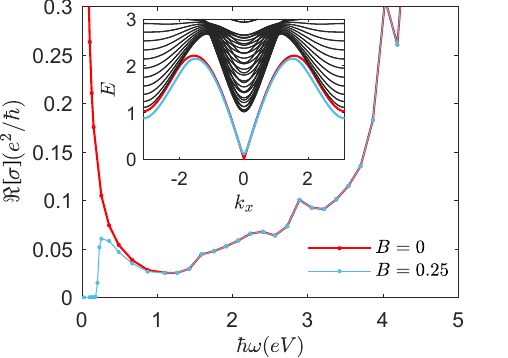}
    \caption{The frequency dependence of the real-part conductivity of the HMMs in the TI thin film model given by Eqs. (\ref{eq:TI})-(\ref{eq:TIb}), with the parameters $A_t=1$, $t_f=1$, $\alpha=1$, $m_0=3$, and $\Delta=2$. The inset is the energy spectrum where the edge states are highlighted by corresponding colors.}
    \label{fig:fig4}
\end{figure}

Adding time-reversal breaking terms will open small gaps at these points. As shown in Fig. \ref{fig:fig4}, the frequency dependence of the real-part conductivity near the $k_x=0$ point (i.e., near $ \omega =0$) has a similar functional form  to that of the $p\pm ip$-wave TSCs and agrees with Eq. (\ref{eq:4}). 

\section{Doped QSH insulator}
\begin{figure}[t]
    \centering
    \subfigure{\includegraphics[scale=1]{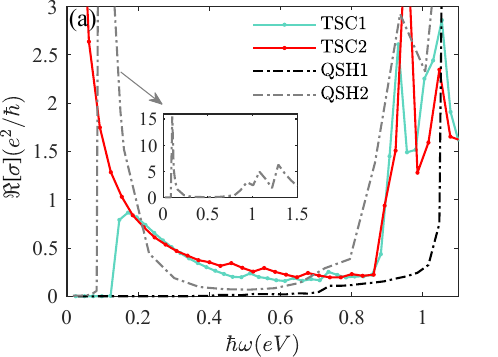}}\vspace{-2mm}
    \subfigure{\includegraphics[scale=0.95]{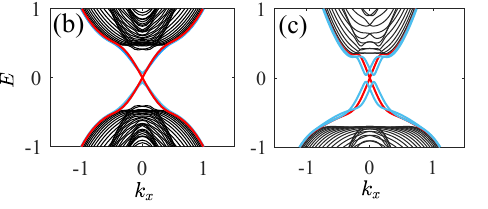}}\vspace{-2mm}
    \caption{(a) The real-part conductivity of the doped QSH system in the QSH phase and in the TSC phase. For the TSC state the parameters are: $A=2$, $\Delta=0.5$, $t=2$, $m_0=0.1$, $A1=2$, $Z=0$ (for the red line) and $Z=0.2$ (for the blue line). For the QSH state: $A=2$, $\Delta=0$, $t=2$, $m_0=0.7$, $A1=2$ and $Z=0$ (for the black line). 
   	(b) and (c) are the energy spectrum of the TSC phase and the QSH phase, respectively, where the time-reversal invariant (broken) edge states are in red (blue). }
    \label{fig:5}
\end{figure}
Quantum spin Hall insulators are proposed to be a TSC through correlation effects \cite{WangPRB2014}. With this model system, one can directly compare the optical response of the HHMs to that of helical normal fermions, which could be achieved in different parameter regimes. 

Consider the following Hamiltonian describing a QSH insulator \cite{BHZ}
\begin{equation}
    {\cal H}_0=M(\bm{k})\sigma_0\tau_3+A(k_x\sigma_3\tau_1-k_y\sigma_0\tau_2),
\end{equation}
where the Pauli matrices $\sigma_i$ and $\tau_i$ ($i=1,2,3$) act on the spin and orbital spaces, respectively. $M(\bm{k})=m_0-t(k_x^2+k_y^2)$ and $m_0t>0$ is required to guarantee the non-trivil topology of the normal state. It has been predicted that a TSC phase with $\Delta_{\mu\nu}^{12}(\bm{k})=\Delta c_{1,\bm{k}\mu}c_{2,-\bm{k}\nu}\delta_{\mu\nu}$ is favored at certain doped region with an inversion-breaking Rashba SOC \cite{WangPRB2014} 
\begin{equation}
    {\cal H}_R=A_1(k_x\sigma_2-k_y\sigma_1)\otimes(\tau_3+\tau_0).
\end{equation} 
Here the subscript $\mu(\nu)$ labels the electron spins and the superscript $1(2)$ represents different orbitals. 

The TSC phase has a pair of helical Majorana states propagating along the boundary, which is replaced by normal-fermion edge states when the pairing term vanishes and  the system transforms into a QSH phase. In presence of a time-reversal breaking term ${\cal H}_Z=Z\sigma_3\tau_0$, the edge states develop new features including a small gap, as shown in Fig. \ref{fig:5}(b) and (c). 

The real-part conductivity of the TSC phase and the QSH phase under uniform detecting light are shown in Fig. 4(a), where the TSC results have similar features to the former TSC models, consistent with Eq. (\ref{eq:4}).  
The results for the QSH phase are rather different, with the optical conductivity inside the topological gap vanishing if $Z=0$. When $Z\neq 0$, it has a sharp peak at $\omega=Z$ (diverging if $T=0$) above which it decreases as $\omega$ goes up. 

The vanishing optical absorption by the QSH edge states forms a major difference from the HMMs. It originates from the different mechanisms through which the edge states couple with electromagnetic waves. While the light-coupling of the HMMs relies on the bulk system and the corresponding current operator must be obtained by projecting the bulk version to the edges, the current operator of the QSH edge states may be directly derived within the effective edge theory which preserves the U(1) gauge symmetry. By introducing a gauge field to the edge theory, the current operator, $j_n\sim v(\psi_1^\dagger\psi_1-\psi_2^\dagger\psi_2)$, can be readily obtained without referring to the bulk. It is simply the $k$-derivative of the edge Hamiltonian $h_{edge}(k)=vk(\psi_1^\dagger \psi_1-\psi_2^\dagger\psi_2)$. 
Following the same procedures in the previous effective 1D theory of Majorana edges states, we get the optical absorption for the QSH edge states at zero temperature, 
$
	\Re[\sigma_\text{QSH}(\omega)]\sim\omega^{-2}(\omega^2-4M^2)^{-1/2},
$
where $M$ is the edge gap open by time-reversal symmetry breaking. Note that, besides the above major difference, the optical responses of the QSH edge states and the HMMs may happen in different ranges of the wavelength since a QSH insulator may have a much larger topological gap.

\section{Conclusion and discussion}  
We have demonstrated that helical Majorana modes induce microwave absorption. It originates from inter-band optical transition processes that are made possible by the broken U(1) gauge and spatial-inversion symmetries. Analytical form of the resulting optical conductivity is obtained with an effective edge theory, which are qualitatively confirmed by numerical calculations with several models of topological superconductors. 
The zero-temperature real-part optical conductivity $\Re[\sigma(\omega)]$ induced by the helical Majorana modes under uniform light is proportional to $\omega^{-1}$. 
When the time-reversal symmetry is broken and an energy gap of $M$ is opened on the edge, $\Re[\sigma(\omega)]$ has a maximum value at $\omega=2\sqrt{2} M$. 
In comparison, vertical optical transitions in helical normal edge states in quantum spin Hall insulators are forbidden unless the time-reversal symmetry is broken after which the functional form of $\Re[\sigma(\omega)]$ becomes similar to that of Majorana modes. This difference originates from the different mechanisms of coupling with the U(1) gauge field. 

Our results show that optical measurements may provide evidence of Majorana edge states in time-reversal invariant topological superconductors. Different from reference \cite{he2021prl}, the detecting light here is uniform and experiments will not encounter the difficulty of focusing a light beam into a tiny spatial region. 
A possible difficulty may come from the background optical absorption signal induced by the bulk Cooper pairs, which may be much larger than the edge-state contribution and make the Majorana signal hard to distinguish. One way to overcome this problem is to tune the external magnetic field which changes the functional form of the Majorana contribution qualitatively while its effect on the background signal is only quantitatively. In this way, it is possible to extract the Majorana contribution.

\section*{Acknowledgments}
We thank Qian Niu and Zhenyu Zhang for helpful discussions. 
J.J.H. is supported by the Innovation Program for Quantum Science and Technology (Grant No. 2021ZD0302800) and the National Natural Science Foundation of China (Grant No. 12204451).

\section*{Appendix: Derivation of the optical conductivity with the effective edge theory}
\renewcommand\theequation{A\arabic{equation}}
\setcounter{equation}{0}
To illustrate the optical response of the HMMs we start from the real space current operator $j(x)=j_a(x)+j_b(x)$ including both intra-band and inter-band current. They can be regarded as a projection of the bulk current onto the edge states. Notice the lowest order of the inter-band current does not contain other terms due to the self-conjugation of Majorana fermions. After the Fourier transformation we have
\begin{align}
    &j_a(q)=a\sum_k k(\gamma_{1,q-k}\gamma_{1,k}+\gamma_{2,q-k}\gamma_{2,k}),\\
    &j_b(q)=-ib\sum_k\gamma_{1,q-k}\gamma_{2,k}.
\end{align}
Define the current-current correlation function $\Pi(q,\tau)=-\langle T_\tau j^\dagger(q,\tau)j(q)\rangle$ we get the corresponding intra-band correlation function

\begin{align}
    \Pi_{a}&=-a^2\sum_{i,k,k^{'}}kk^{'}\langle T_\tau [\gamma_{i,-k-q}\gamma_{i,k}]_\tau[\gamma_{i,q-k^{'}}\gamma_{i,k^{'}}]\rangle\\ 
    &=-a^2\sum_{i,k}(2k^2-kq)\mathcal{G}_i(-k,\tau)\mathcal{G}_i(k-q,\tau).
\end{align}
where $i=1,2$ and $\mathcal{G}_i$ stands for the Green's function of $\gamma_1$ and $\gamma_2$. In the frequency space at $T\rightarrow 0$ limit we have
\begin{align}
    \Pi_a(i\omega_n)=&\int_{0}^{k_BT}\Pi_a(q,\tau)e^{i\omega_n\tau}
    \\=&-a^2\sum_k(2k^2-kq)[\theta(k)-\theta(k-q)]\notag \\&\times (\frac{1}{i\omega_n+vq}+\frac{1}{i\omega_n-vq}).
\end{align}
The $q=0$ current operator is given by
\begin{align}
    j(k)=a k\sigma_0+b\sigma_3
\end{align}
under the basis $\Gamma_k^\dagger=[\gamma^\dagger_{1,k},\gamma^\dagger_{2,k}]$. Under the Bogoliubov transformation we can diagonalize the 1D Hamiltonian $H\rightarrow\tilde{H}=UHU^\dagger,\quad \Gamma^\dagger\rightarrow\tilde{\Gamma}^\dagger=[\tilde{\gamma}_1^\dagger,\tilde{\gamma}_2^\dagger]$. The current operator should become
\begin{align}
    \tilde{j}(k)&=a k\sigma_0+b \begin{pmatrix}u_k&v_k\\v_k^*&-u_k^*\end{pmatrix}  \sigma_2
       \begin{pmatrix}u_k^*&v_k\\v_k^*&-u_k\end{pmatrix}\\
    &=a k\sigma_0+b\begin{pmatrix} i \Im (v_ku_k^*) & i(u_k^2+v_k^2)\\  -i(u_k^{*2}+v_k^{*2}) & -i \Im (u_k^*v_k)\end{pmatrix}
    \label{eq:A13}
\end{align}
in the new basis. To maintain the Majornana algebra of the new quasi-particle states $\{\tilde{\gamma}_{ik},\tilde{\gamma}_{jk^{'}}\}=\delta_{ij}\delta_{k,-k^{'}}$ the parameters $u_k=\vert u_k\vert e^{i\phi_u}$ $v_k=\vert v_k\vert e^{i\phi_v}$ should satisfy
$
    \vert u_k\vert^2=\frac{1}{2}+\frac{1}{2}\frac{\epsilon_k}{\xi_k},
    \vert v_k\vert^2=\frac{1}{2}-\frac{1}{2}\frac{\epsilon_k}{\xi_k}$, and $
    \phi_u-\phi_v=\frac{\pi}{2}.
$

The $a k \sigma_0$ term from Eq. (\ref{eq:A13}) does not contribute to the correlation function because it only involves terms like $\langle T_\tau \tilde{\gamma}_i^\dagger(\tau)\tilde{\gamma}_i(\tau)\tilde{\gamma}_i^\dagger\tilde{\gamma}_i\rangle$ ($i=1,2$). Such terms vanish under Wick's theorem since the involved Green's functions belong to the same Majorana operator. However the second term of Eq. (\ref{eq:A13}) will cause connected diagrams  
\begin{align}
\Pi_{21}(\tau)=-4b^2\sum_{k>0}\vert u_k\vert^2\vert v_k\vert^2\langle T_\tau \tilde{\gamma}_{2k}^\dagger(\tau)\tilde{\gamma}_{1k}(\tau)\tilde{\gamma}_{1k}^\dagger\tilde{\gamma}_{2k}\rangle
\end{align}
corresponded to the inter-band transition. In the frequency space we have
\begin{align}
\begin{aligned}
\Pi_{21}(i\omega_n)=-b^2\sum_{k>0}\frac{\epsilon_k^2}{\xi_k^2}\frac{f(\xi_k)-f(-\xi_k)}{i\omega_n-2\xi_k}.
\end{aligned}
\end{align}
And the real-part optical conductivity is given by
\begin{align}
    \Re[\sigma(\omega)]&=-\frac{1}{\omega L}\Im[ \Pi_{21}(i\omega_n)],\\
    &=\frac{b^2}{2v\omega^2}{\sqrt{\omega^2/4-M^2}}\tanh{\frac{\omega}{2k_BT}}.
    \label{eq:A17}
\end{align}
\email{jun\_he@ustc.edu.cn}
\bibliography{ref.bib}

\end{document}